# A Nonparametric Approach to Measure the Heterogeneous Spatial Association: Under Spatio-Temporal Data

*Zihao Yuan*[1]

Spatial association and heterogeneity are two critical areas in spatial analysis, geography, statistics and so on. Though large amounts of outstanding methods have been proposed and studied, there are few of them tend to study spatial association under heterogeneous environment. Additionally, most of the traditional methods are based on distance statistic and spatial weighted matrix. However, in some abstract spatial situations, distance statistic cannot be applied since we cannot even observe the geographical locations directly. Meanwhile, under these circumstances, due to the invisibility of spatial positions, designing of weight matrix cannot absolutely avoid subjectivity. In this paper, a new entropy-based method, which is data-driven and distribution-free, has been proposed to help us investigate spatial association while fully taking the fact that heterogeneity widely exists. Specifically, this method is not bounded with distance statistic or weight matrix. Asymmetrical dependence is adopted to reflect the heterogeneity in spatial association for each individual and the whole discussion in this paper is performed on spatio-temporal data with only assuming stationary m-dependent over time.

Keywords: spatial association; heterogeneity; asymmetrical dependence; entropy

## 1. Introduction

### *1.1 Statement of Problems*

Spatial association (or spatial dependence) has been widely researched by statisticians with the development of relative subjects, such as spatial statistics, geography, regional sciences and so on. Many scholars offered us nice tools to investigate the existence, structure and power of spatial association. The most celebrated tools are Moran's I, Geary's C and Local Indicator of Spatial Association (LISA). Though Moran's index

[1] Department of Statistics, School of Data Science, Zhejiang University of Finance and Economics, Hangzhou, China

could tell us whether there exists spatial association, the index only demonstrates it globally. On the other hand, by absorbing the framework of Moran's I, Geary's C helps us to check more information about local spatial association. Anselin (1995) generalized the idea "local", he discussed a class of statistics (indicators) used to dredge up information about local spatial association. More specifically, he demonstrated how these indicators can make contributions to each individual in the neighborhood. However, all the three masterworks depend on the design of weight matrix. It is well known that, in some abstract spatial situations, designing spatial weight matrix needs to count on the background of the problems we investigate, which, in other words, requires some critical empirical knowledge about the topic. No doubt, when the weight matrix perfectly reflects the dependence among individuals, the result would be nice. Unfortunately, it cannot always be considerably done. By the way, we humbly suppose that since we want to use techniques from data science, we should let data "speak" for itself to avoid unnecessary subjectivity.

Another critical point is, except for spatial association, spatial heterogeneity is also a common phenomenon we should not ignore. There are several different aspects of heterogeneity frequently used in geography and sociology. Two most important ones are spatial local heterogeneity and spatial stratified heterogeneity. The first one shows that, given a measurement, its value varies at different positions, while the second one presents the varying values of measurement in different hierarchies. Actually, in social science, stratified heterogeneity can sometimes also be transformed into local heterogeneity, since the stratification here can somehow be a generally-defined abstract location. For example, when we are investigating the heterogeneity of every student's academic capability in the same class, traditional geographical position does not suit this situation. We normally tend to stratify these students by their grades or some other indexes. But please note that grades itself, no matter points or ranking, they are actually some kinds of abstract positions with natural ordering. Hence, motivated by this example, in this paper, once we mention heterogeneity, it means local heterogeneity of each individual bounded up with a spatial position, no matter whether they are geographical, abstract or even unrecognizable.

Unfortunately, association and heterogeneity, these two properties are often involved in each other. For example, suppose there are N enterprises selling the same kind of goods with different quality on line. Owing to different technical strength and financial status, each enterprise delivers different effects to any other one in this market, particularly on pricing strategy. Obviously, technological distinction, financial strength or some other resources can be seen as a representation of heterogeneity. Considering the effect delivered by each individual here is dissimilar, we are allowed to say the association between any two enterprises is asymmetrical. On the other hand, asymmetries in spatial association inversely make spatial heterogeneity more significant. Thus, in this paper, we attempt to use asymmetrical dependence to represent heterogeneous spatial association for each individual and the main purpose of this article contains two points: (a)under spatial-temporal data(defined as AS.1), we build a general method, which is data-driven, distribution-free and independent of weight matrix and geographical positions, to measure the heterogeneous spatial association particularly for each individual; (b) based on the method of (a), we tend to discuss the asymptotic properties of it, especially consistency, convergence rate and asymptotic normality.

### *1.2 Previous work review*

Measuring the heterogeneous spatial association is a relatively new area in regional science, especially in spatial data analysis. Based on the limited literature we can search, though LISA is an acceptable method, Li (2008) might be the first one who proposed a direct approach being independent of weight matrix to investigate the spatial association under heterogeneous environment. Li used scaleable moving window (SMW) technique to measure the spatial association. A key point in this article is that, for each window, this algorithm will produce a set of association rules within the locations investigated, where the spatial association pattern is represented by support value and confidence value calculated by Apriori algorithm. Then, Sha & Li (2009) offered another method to this topic. They tended to calculate the association between individual and its neighbor one by one, in the light of the distribution of spatial locations and individuals interested. Generally speaking, both of the two approaches deal with spatial association locally, which suggests the asymmetries in dependence are reflected. After all, for a spatial process, heterogeneity is always bounded with non-stationarity. Notwithstanding these methods are effective, they still rely overly on geographical distance, which cannot fit the main purpose of this paper. On the other hand, we still need to specify the representation of the heterogeneity of spatial association. Please note for the arbitrary i-th individual, if we know exactly which ones are important to it, we can precisely capture the heterogeneity involved in spatial association. According to the conclusion of idea from the two techniques mentioned above, a natural idea is to apply asymmetrical dependence to illustrate the heterogeneity of spatial association.

As for asymmetrical dependence estimation, it has been widely studied in time series analysis, finance and portfolio management. Conditional correlation proposed by Ang & Chen (2007), Hong, Tu & Zhou (2008) (hereafter Hong) might be the earliest research on testing asymmetrical dependence. Specifically, Hong's method firstly offered us a model-free test of asymmetric exceedance correlation. The null hypothesis is as follow,

$$H_0 : \rho^+ = \rho^-,$$

where
$$\rho^+ = Corr(X_{1t}, X_{2t} \mid X_{1t} > \mu_1 + c\sigma_1, X_{2t} > \mu_2 + c\sigma_2),$$

$$\rho^- = Corr(X_{1t}, X_{2t} \mid X_{1t} < \mu_1 - c\sigma_1, X_{2t} < \mu_2 - c\sigma_2),$$

and $\{X_{1t}\}$ , $\{X_{2t}\}$ are two stationary time series. By the way, $\mu$ and $\sigma$ indicate expectation and variance respectively. Besides, under the null hypothesis, Hong et al. discovered the asymptotic distribution of this test statistic, $J = (\rho^+ - \rho^-)^{\mathrm{T}} \Omega^{-1} (\rho^+ - \rho^-)$, is Chi-square distribution. However, since this method only absorbed "linear" dependence, Embrechts, McNail & Straumann (2002) discovered that, when we applied this method to stock return research, linear exceedance correlation could not fully capture the whole information of dependence. Skuang & Tjøstheim (1993), Maasoumi & Racine (2008) showed us theoretical properties of using mutual information to identify the dependence. More recently, Manner(2010) developed a method to test of exceedance correlation based on copula function. Giannerini, Maasoumi & Dagum (2015) discussed the asymptotic distribution of a class of general statistics of dependence. The latest literature about the test of asymmetrical dependence is Jiang et al. (2017) (hereafter Jiang). They combined mutual information and exceedance correlation together to measure the tail dependence. Meanwhile, they also widely investigated some simulation-based finite sample properties and asymptotic size of the test. Unlike Jiang's work which is based on simulation work, we prefer to concentrate solely on theoretical properties of the method proposed in this paper. Besides, please recall that, our main goal is to measure the spatial association under heterogeneous environment. Thus, we cannot make any assumption of spatial process, like stationary. Meanwhile, research on the test of asymmetrical dependence in time series analysis has been very fruitful. Considering that most of spatio-temporal data can be viewed as a collection of N time series with spatial association, our work is all based on spatio-temporal data, which is non-stationary over space while stationary m-dependent over time.

Another challenge here is to estimate the mutual information defined as follow,

$$I(X_1,X_2) \overset{def}{=} I_{12} = \int_S f_{12}(x_1,x_2)\log\frac{f_{12}(x_1,x_2)}{f_1(x_1)f_2(x_2)}dx_1dx_2\,, \tag{1-1}$$

$S$ here is the support set. $X_1$ and $X_2$ are random variables (vectors). $f_i$ and $f_{ij}$ denote marginal and joint density function respectively. Since we tend to build a model-free method, we can only use nonparametric approach. Fortunately, according to Shannon's information theory, we can decompose the mutual information as follow[2],

$$I(X_1,X_2) = H(X_1) + H(X_2) - H(X_1,X_2)\,, \tag{1-2}$$

where for any random variable(vector) $X$,

$$H(X) = -\int_S f(x)\log f(x)dx\,, \tag{1-3}$$

is defined as the entropy of $X$. Therefore, we only need to obtain a consistent nonparametric estimator of entropy.

In this area, there has been whopping pile of outstanding previous work. The most prevalent idea of nonparametric entropy estimation is "plug-in", which means we estimate the density at first and plug it into some functional estimator. Please note that, based on (1-3), it is obvious that entropy is essentially same as $-E(\log f(x))$. Hence, a natural idea is to use sample mean of random variable log $f(x)$ to estimate it, indicating the entropy estimator can be expressed as follow,

$$\hat{H}(X) = -\frac{1}{n}\sum_{i=1}^{n}\log\hat{f}(x_i)\,. \tag{1-4}$$

[2] Another feasible approach is to make use of copula function since $\log C(u,v) = \log\frac{f_{12}}{f_1 f_2}$, where u=$F_1$, v=$F_2$. This connection will be discussed in our supplementary materials.

There is no doubt that histogram or kernel-based entropy estimators are two of the most widely-researched methods. Beirlant et al.(2001) wrote a nice literature review of entropy estimation. Among these works, one of the most significant ideas proposed by Györfi & van der Meulen (1987, 1989, 1990)(hereafter GM) is splitting-data method. Here we decompose a sample, $\{X_1,\ldots,X_n\}$, into two sub-samples, $\{X_1,\ldots,X_k\}$ and $\{Z_1,\ldots,Z_m\}$, where $m+k=n$. We firstly apply $\{X_1,\ldots,X_k\}$ to construct density estimation, then use this estimation and the second sub-sample to estimate the entropy. Its estimator can be written as follow,

$$\hat{H} = -\frac{1}{m}\sum_{i=1}^{m}\log \hat{f}(Z_i)I\{\hat{f}(Z_i) > a_k\}, \tag{1-5}$$

where $a_k$ satisfies $\lim_{k\to+\infty} a_k = 0$ . More specifically, $\hat{f}$ was considered as histogram, kernel or any $L_1$-consistent density estimator in 1987, 1989 and 1990 respectively. A prominent advantage of this method is that, it can reduce the complexity of calculation and require only mild tail and smoothness conditions. However, this work and the others mentioned in that review are still based on independent and identically distribution (hereafter i.i.d.) sample. About dependent data, Lim (2007) proved a strong consistency of GM under φ-mixing sequence. More recently, Källberg et al. (2014) proposed a Rényi entropy estimator based on U-statistic under stationary m-dependent data. In this article, our entropy estimator shares the same idea, "splitting-data", as (1-5).

## 2. Estimation

In this section, we are going to discuss the construction and asymptotic properties of the estimator built to measure the heterogeneous spatial association. At first, we need to present the following assumptions necessary for our theoretical discussion.

**Assumptions**

**AS.1** The spatial lattice of the spatio-temporal process is regular, i.e. this process can be written as $\{X(\mathbf{i},t)|\mathbf{i}\in\mathbf{Z}^N, t\in R^+\}$ and $X(\mathbf{i},t)\in R^1$, where N is a positive integer and $\mathbf{Z}$ denotes the set of whole integers.

**AS.2** The spatio-temporal data over time is stationary m-dependent and satisfies the following equality, $\sup_t \sup_{A,B\in M|t-s|\le m}\left|\frac{Cov(A,B)}{\sqrt{Var(A)}\sqrt{Var(B)}}\right| = \rho(m)$, where m is a fixed positive integer and t, s are positive real numbers. Here $M_{|t-s|\leqq m}$ denotes the $\sigma$ field of any limited-dimensional random sequence ($X$(t),...,$X$(s)), where |t-s|≦m and $X$(t) here could be a univariate or bivariate vector. Besides, we deliver no assumption to the data over space.

**AS.3** For any marginal density function $f_i$, and joint density of two individuals $f_{ij}$, the following conditions are satisfied:

(1) They are square integrable functions and larger than 0.

(2) They have third derivatives and continuous second derivatives written as $f_i^{''}, f_{ij}^{''}$.

(3) These integrals depicted below exist on support sets,

$\int f_i^{''}(x)f_i(x)dx$, $\int f_{ij}^{''}(x)f_{ij}(x)dx$, $\int f_i^{''}(x)f_{ij}(x)dx$, $\int (f_i^{''}(x))^2 f_i(x)dx$, $\int (f_{ij}^{''}(x))^2 f_{ij}(x)dx$.

**AS.4** For any marginal entropy $H_i$, and joint entropy $H_{ij}$, they are finite and positive.

**AS.5** Variance and covariance of random variables $\log f_i$, $\log f_{ij}$ exist.

**AS.6** The kernel functions we choose to estimate density functions should satisfy the following conditions:

(1) $K(v)$ is a positive and symmetric function.

(2) $\int K(v)dv = 1$, $\int vK(v)dv = 0$, $\int v^2K(v)dv < +\infty$, $\int (K(v))^2 dv < +\infty$.

**AS.7** For any given i, we have $h_i \to 0$, $T \to +\infty$, $[\frac{T+1}{2}]h_i h_j \to +\infty$, $\sqrt{[\frac{T+1}{2}]}h_i h_j \to 0$ and $h_i = O(h_j)$ ,where $h_i$ and $h_j$ denote bandwith chosen for smoothing density functions $f_i$ and $f_j$ respectively.

**AS.8** We say individual (location) **j** is important to **i** if their association is vital to **i**.

Please note the last assumption (AS.8) actually serves as the foundation of discussion about heterogeneous spatial association. It is reasonable because **i** can be affected by **j** only if they are associated with each other. Besides the information delivered from **j** to **i** is totally contained by mutual information.

Remark that owing to that index set $\{\mathbf{i}|\mathbf{i} \in \mathbf{Z}^N\}$ is a countable set, we are allowed to replace each vector i with a simple scalar i. Actually, this would make no difference to the result of our estimation and the following two points can explain the reason for that: (1)as we discussed before, we try to represent the heterogeneity involved in spatial dependence by "locally, relatively and asymmetrically" spatial association, hence for every spatial individual i, spatial association will be remeasured; (2)since we might be confronted with some kinds of spatial situation where locations are invisible, vector **i** is actually unknown, which means we have to attach a label to each location or individual and the property —countable— allows us to do it.

### *2.1 Construction of Estimators*

Under AS.4, via (1-2) and information theory, we can easily demonstrate the relationship between mutual information and entropy by Venn's diagram (see Fig. 1),

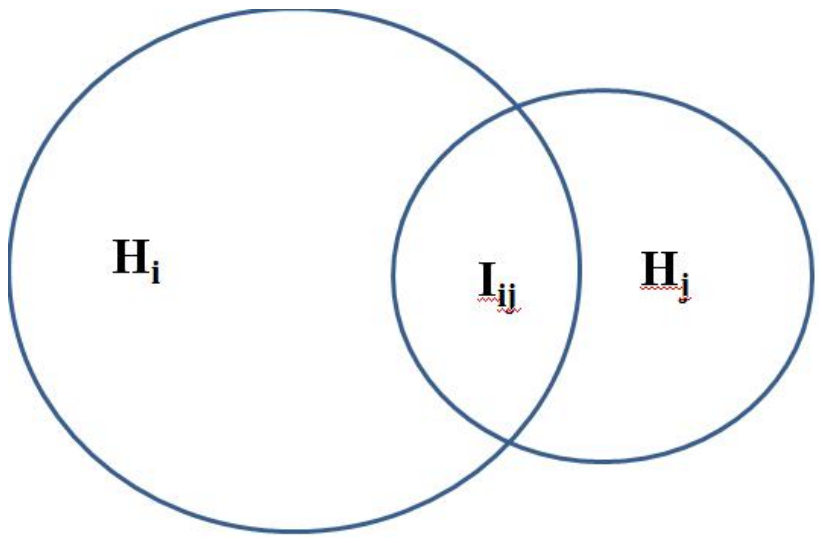

Figure 1. Venn's diagram of the relationship between mutual information and entropy

In the Figure 1, $I_{ij}$ and $H_i$ (or $H_j$) denote the mutual information and entropy respectively. Please note that $I_{ij}$ is larger than 0 if and only if $X_{it}$ is associated with $X_{jt}$, for any given t. More precisely, according to AS.2, we can say, $I_{ij}=0 \Leftrightarrow X_i$ is independent from $X_j$. Please note the mutual information helps us to quantify the association between spatial individual (or some invisible location) i and j. Meanwhile, $H_i$ and $H_j$ are generally unequal to each other since heterogeneity exists. Thus, according to AS.6, a natural idea is to use the ratio between mutual information and entropy to reflect how important j is to i, written as follow,

$$r_{ij} = \frac{I_{ij}}{H_i}. \tag{2-1}$$

Ulteriorly, it naturally has the following properties:

(1) $r_{ij}=r_{ji}$ is correct if and only if $H_i=H_j$.

(2) $0 \le r_{ij} \le 1$, since for any i and j, $I_{ij} \le H_i$.

(3) $r_{ij}=0$ if and only if $I_{ij}=0$ ,which means individual i is independent from individual j.

These properties can be simply proved by using properties of entropy so we skip it here. Please recall (1-2), estimation of $r_{ij}$ is allowed to be transformed into estimation of entropy. Therefore, our estimator is expressed as follow,

$$\hat{r}_{ij} = \frac{\hat{I}_{ij}}{\hat{H}_i} = 1 + \frac{\hat{H}_j - \hat{H}_{ij}}{\hat{H}_i}. \tag{2-2}$$

As we discussed in section 1.2, we absorb the idea of splitting-data method. Hence,to estimate $H_{ij}$, we decompose its sample sequence $\{X_i(t), X_j(t)|t \in R^+\}$ into odd and even sub-sequences written as $\{Z_{i1},...,Z_{i[(T+1)/2]}, Z_{j1},....Z_{j[(T+1/2]}\}$ and $\{X_{i1},...,X_{i[T/2]}, X_{j1},....X_{j[T/2]}\}$ respectively, where T is the sample size of time and [R] denotes the integer part of any real number R. We firstly apply odd sub-sequence to construct the kernel-based density estimator, and then by using even sub-sequence and density estimation, we get the estimation of entropy $H_{ij}$. Under this procedure, our entropy estimator is expressed as follow,

$$\hat{H}_{ij} = -\frac{1}{[\frac{T}{2}]}\sum_{t=1}^{[\frac{T}{2}]}\log \hat{f}_{ij}(X_{it}, X_{jt}) \overset{def}{=} -\frac{1}{[\frac{T}{2}]}\sum_{t=1}^{[\frac{T}{2}]}\log \hat{f}_{ij}\,, \qquad (2\text{-}3)$$

where

$$\hat{f}_{ij}(x_i, x_j) = \frac{1}{[\frac{T+1}{2}]h_i h_j}\sum_{t=1}^{[\frac{T+1}{2}]} K_i\left(\frac{Z_{it} - x_i}{h_i}\right) K_j\left(\frac{Z_{jt} - x_j}{h_j}\right). \qquad (2\text{-}4)$$

Similar to the joint entropy, this construction works the same to marginal entropy. The only difference here is that we replace product-kernel with single kernel, so we have,

$$\hat{H}_i = -\frac{1}{[\frac{T}{2}]}\sum_{t=1}^{[\frac{T}{2}]}\log \hat{f}_i\,, \qquad (2\text{-}5)$$

where

$$\hat{f}_i(x_i) = \frac{1}{[\frac{T+1}{2}]h_i}\sum_{t=1}^{[\frac{T+1}{2}]} K_i\left(\frac{Z_{it} - x_i}{h_i}\right). \qquad (2\text{-}6)$$

### *2.2 Consistency*

In this part, we mainly investigate the consistency and convergence rate of $r_{ij}$. Obviously, if we investigate the consistency of our estimator $\hat{r}_{ij}$ , according to Slutsky theorem, we only need to obtain the consistency of entropy estimator. Without loss of generality, we here discuss convergence of estimator of $H(X_1, X_2)$ abbreviated to$H_{12}$, where $(X_1,X_2)$ is a two-dimensional random vector. At first, there are some preliminary lemmas and corollaries which are necessary to our discussion.

**Lemma 1**.*Under assumption AS.2-AS.7 and (2-4), for any given ($x_1$, $x_2$) from support set of ($X_1$,$X_2$), $\hat{f}_{12}(x_1, x_2)$ is supposed to be a mean-square consistent estimator of $f_{12}(x_1, x_2)$ with the following results,*

$$Bias(\hat{f}_{12}(x)) = \frac{1}{2}\left[\sum_{l=1}^{2}\left((\kappa_{2(l)})h_l\right)^2 f_{ll}''(x)\right] + O(\sum_{l=1}^{2} h_l^3) \ , \tag{2-7}$$

$$Var(\hat{f}_{12}(x)) \le \left(\frac{2[\frac{m}{2}]\rho([\frac{m}{2}])+1}{[\frac{T+1}{2}]h_1h_2}\right)\left(f_{12}(x)\int\prod_{l=1}^{2} K_l^2(v)dv + O(\sum_{l=1}^{2} h_l^2+1)\right), \tag{2-8}$$

$$\hat{f}(x) = O_p\left((\sum_{l=1}^{2} h_l^2)^2 + ([\tfrac{T+1}{2}]h_1h_2)^{-\frac{1}{2}}\right), \tag{2-9}$$

where $\kappa_{2(l)} = \int v^2 K_l(v)dv$ , $l = 1,2$ and $x = (x_1, x_2)$ . Apparently, (2-9) is just a natural extension of (2-8) and (2-7). Though proof of lemma 1 is simple and tedious, considering the completeness of this article, we demonstrate it in supplementary materials. Similar to lemma 1, for any marginal density, the following corollary 1 presents some asymptotic properties of marginal density estimator,

**Corollary 1**. *For random variable $X_i$ of individual i, under AS.2-AS.7 and (2-6), we have the following results,*

$$Bias(\hat{f}_i(x_i)) = \frac{(\kappa_{2(i)})^2}{2}\left[h_i^2 f_i''(x_i)\right] + O(h_i^3), \tag{2-10}$$

$$Var(\hat{f}_i(x_i)) \le \left(\frac{2[\frac{m}{2}]\rho([\frac{m}{2}])+1}{[\frac{T+1}{2}]h_i}\right)\left(f_i(x_i)\int K^2(v)dv + O(h_i^2)\right), \tag{2-11}$$

$$\hat{f}_i(x_i) = O_p\left(h_i^2 + ([\tfrac{T+1}{2}]h_i)^{-\frac{1}{2}}\right). \tag{2-12}$$

Since corollary 1 is a special case of lemma 1, so we skip its proof here. Based on lemma 1, mean-square consistency of our entropy estimator written as $\hat{H}_{12} = -\frac{1}{[\frac{T}{2}]}\sum_{t=1}^{[\frac{T}{2}]}\log\hat{f}_{12}(x_{1t}, x_{2t})$, can be obtained, shown as following lemma 2.

**Lemma 2**. *According to lemma 1, AS.2-AS.7 and $\hat{H}_{12}$, we can obtain following results,*

$$\hat{H}_{12} - H_{12} = O_p\left(\left(\sum_{l=1}^{2} h_l^2\right)^2 + \left[\frac{T}{2}\right]^{-\frac{1}{2}}\right), \tag{2-13}$$

$$Bias(\hat{H}_{12}) = \frac{1}{2\zeta}\left[\sum_{l=1}^{2}\left(\kappa_{2(l)}h_l\right)^2\int f_{ll}''(x)f(x)dx\right] + O(\sum_{l=1}^{2}h_l^3), \tag{2-14}$$

$$Var(\hat{H}_{12}) = [\tfrac{T}{2}]^{-1}Var(\log f_{12}(x)) + o(1), \tag{2-15}$$

*where* $\zeta = \theta\hat{f}_{12}(x) + (1-\theta)f_{12}(x)$ *and* $\lim_{T\to+\infty}\zeta = f_{12}(x)$ *, for any given* $x = (x_1, x_2)$ *from support set of* $(X_1,X_2)$*.*

**Theorem 1**. *According to lemma 2, under AS.2-AS.7, we have*

$$\hat{r}_{ij} \xrightarrow{p} r_{ij} \tag{2-16}$$

No doubt, by applying lemma 2 and Slutsky's lemma, theorem 1 gets proved immediately (see proof of lemma 2 in supplementary materials). In addition, we can also obtain the properties of mean-square consistency for marginal entropy of any i-th individual, $H_i$, by using the technique from the proof of lemma 2. Thereby we just enumerate it here as corollary 2 without proof.

***Corollary 2****. Under Corollary 1, AS.2-AS.7 and (2-5), we directly have*

$$\hat{H}_i - H_i = O_p(h_i^4 + [\tfrac{T}{2}]^{-\frac{1}{2}}), \tag{2-17}$$

$$Bias(\hat{H}_i) = \frac{\left(\kappa_{2(i)}\right)^2}{2\eta_i}\left[h_i^2\int f_i''(x_i)f_i(x_i)dx_i\right] + O(h_i^3), \tag{2-18}$$

$$Var(\hat{H}_i) = [\tfrac{T}{2}]^{-1}Var(\log f_i(x_i)) + o(1), \tag{2-19}$$

*where* $\eta_i$ *is an intermediate value between* $\hat{f}_i(x_i)$ *and* $f_i(x_i)$ *converging to* $f_i(x_i)$ *as* $T \to +\infty$*, for any x from support set of* $X_i$*.*

Another point deserving attention is that, under AS.2, we know the consistency of $\hat{r}_{ij}$ heavily relies on the stationarity and dependence of data over time. Thus, necessity of obtaining the convergence rate and asymptotic bias of $\hat{r}_{ij}$ is significant since sample size is always limited. For that, we regard $\hat{r}_{ij}$ as $\hat{r}_{ij}(\hat{H}_i, \hat{H}_j, \hat{H}_{ij})$ and use the second-order

Taylor's expansion at $(\hat{H}_i, \hat{H}_j, \hat{H}_{ij}) = (H_i, H_j, H_{ij})$. However, in order to fully capture every term in the expansion, the following lemma is required which is also critical in the discussion of approximate normality.

**Lemma 3**. *Under AS.2-AS.7, (2-7), (2-10), (2-14) and (2-18), for any i and j, the following results can be shown (See proof of lemma 3 in Appendix),*

$$E(\hat{H}_i \hat{H}_j) = E(\log f_i \log f_j) + O(h_i^2) + O(h_j^2), \tag{2-20}$$

$$E(\hat{H}_i \hat{H}_{ij}) = E(\log f_i \log f_{ij}) + O(h_i^2) + O(\sum_{l=\{i,j\}} h_l^2) \tag{2-21}$$

Then, according to lemma 3, the following theorem 2 can be shown easily, which illustrates the convergence rate and asymptotic bias of estimation.

**Theorem 2**. *Under AS.2-AS.7 and all the lemmas or corollaries mentioned before, we obtain the convergence rate of estimator $\hat{r}_{ij}$ as follow,*

$$\hat{r}_{ij} - r_{ij} = O_p(h_i^4 + h_j^4 + [\tfrac{T}{2}]^{-\frac{1}{2}}), \tag{2-24}$$

$$Bias(\hat{r}_{ij}) = r_{ij} + O_p(h_i^2 + h_j^2 + \left[\tfrac{T}{2}\right]^{-\frac{1}{2}}). \tag{2-25}$$

Proof of theorem 2 can be seen in supplementary materials.

***2.3 Approximate Normality***

This portion states the approximate normality of estimator $\hat{r}_{ij}$ , which offers us a foundation of discussion in section 2.4. Recall that we consider $\hat{r}_{ij}$ as a function of $(\hat{H}_i, \hat{H}_j, \hat{H}_{ij})$ and this function obviously has continuous first-order preference derivatives, a natural idea is to obtain the asymptotic distribution by delta method which is a well-known technique in large sample theory. For doing that, we firstly investigate the asymptotic distribution of vector $(\hat{H}_i, \hat{H}_j, \hat{H}_{ij})^{\mathrm{T}}$.

**Lemma 4**. *By using lemma 3, the asymptotic covariances between any two of* $\hat{H}_i$, $\hat{H}_j$ *and* $\hat{H}_{ij}$ *are as follow,*

$$Cov(\hat{H}_i,\hat{H}_j) = Cov(\log f_i, \log f_j) + O(h_i^2 + h_j^2), \tag{2-22}$$

$$Cov(\hat{H}_i,\hat{H}_{ij}) = Cov(\log f_i, \log f_{ij}) + O(h_i^2 + \sum_{l=\{i,j\}} h_l^2). \tag{2-23}$$

Through lemma 4, (2-15) and (2-19), we can obtain the covariance matrix of $(\hat{H}_i,\hat{H}_j,\hat{H}_{ij})^{\mathrm{T}}$ and by applying central limit theorem of stationary m-dependent sequence, we instantly get lemma 5.

**Lemma 5**. *Under assumptions AS.2-AS.7, by applying lemma 2, corollary 2, lemma 4 and setting* $(\hat{H}_i,\hat{H}_j,\hat{H}_{ij})$ *as* $\hat{\mathbf{H}}$, $(H_i,H_j,H_{ij})$ *as* $\mathbf{H}$, *we have*

$$\sqrt{[\tfrac{T}{2}]}(\hat{\mathbf{H}}^{\mathrm{T}} - \mathbf{H}^{\mathrm{T}}) \xrightarrow{d} N(0,\Sigma), \tag{2-24}$$

*where*

$$\Sigma = \begin{pmatrix} Var(\log f_i) & * & * \\ Cov(\log f_i, \log f_j) & Var(\log f_j) & * \\ Cov(\log f_{ij}, \log f_i) & Cov(\log f_{ij}, \log f_j) & Var(\log f_{ij}) \end{pmatrix}.$$

According to the lemma 5(see proof in supplementary materials), by using delta method, we finally capture the approximate normality of $\hat{r}_{ij}$ as follow,

**Theorem 3**. *Based on AS.2-AS.7 and all of the lemmas and corollaries above, we have*

$$\sqrt{[\tfrac{T}{2}]}(\hat{r}_{ij} - r_{ij}) \xrightarrow{d} N(0, \nabla_{\mathrm{r}}(\mathbf{H})\boldsymbol{\Sigma}\nabla_{\mathrm{r}}^{\mathrm{T}}(\mathbf{H})), \tag{2-25}$$

*where* $\nabla_r(\mathbf{H})$ *is the gradient of function* $\hat{r}_{ij}$, *when* $\hat{\mathbf{H}} = \mathbf{H}$.

Proof of theorem 3 is just a simple application of central limit theorem and delta method, thus we skip it here.

### *2.4 Significant Test of $r_{ij}$*

Recall that $r_{ij}$ is built to measure how much effect delivered from j to i or how important j is to i and its domain zone is [0,1]. Therefore, we can say, when j is significantly important to i, value of $r_{ij}$ must be closed to 1. More precisely, there exists a real number $r \in (0,1)$, satisfying $r_{ij} \in (c,1)$. According to this property, we set the null hypothesis of our significant test as, $H_0$: $r_{ij} \geq r$. Meanwhile, since in section 2, weak consistency of $\hat{r}_{ij}$ has been proved. Then, under $H_0$, it is obvious that the rejection area should be $\mathrm{W} = \{\hat{r}_{ij} \leq \mathrm{C} \mid \mathbf{X}\}$, where **X** denotes the data set and C is a critical value. Fortunately, by using (2-25), we can calculate C when sample size T is sufficiently large. It is interesting to know, law of large number still exists under stationary m-dependent sequence. Under this circumstance, estimation of asymptotic covariance matrix can be done by simply using sample moments to replace each element in that matrix, written as $\hat{\Sigma}$. And we use $\nabla(\hat{\mathbf{H}})$ to replace $\nabla(\mathbf{H})$, since the gradient here is still a continuous mapping. Thus, we use $\mathbf{S} = \nabla(\hat{\mathbf{H}})\hat{\Sigma}\left(\nabla(\hat{\mathbf{H}})\right)^{\mathrm{T}}$ to represent covariance matrix for calculating C.

Furthermore, according to the definition of criteria level α, the following equality can be obtained,

$$\sup \alpha(r_{ij}) = P\{\hat{r}_{ij} \leq C \mid r_{ij} \geq r\} \leq \alpha, \quad (2\text{-}26)$$

Based on (2-25), when T is large enough, we directly have

$$P\{\hat{r}_{ij} \leq C \mid r_{ij} \geq r\} = \Phi\left(\frac{C - r_{ij}}{\mathbf{S}/[\frac{T}{2}]}\right). \quad (2\text{-}27)$$

Apparently, Φ is a monotonous decreasing of $r_{ij}$. By setting $r_{ij}= r$ and $\alpha = \Phi(\frac{C-r}{\mathbf{S}/[\frac{T}{2}]})$,

our critical value can be expressed as $C = r^* + q_\alpha(\mathbf{S}/[\frac{T}{2}])$, where $q_\alpha$ denotes the quantile of order αof standard normal distribution.

Furthermore, to build the test of asymmetrical dependence between two individuals, we firstly discuss the asymmetrical dependence between two individuals. According to Figure 1 and properties of $r_{ij}$, it is obvious, when $r_{ij} = r_{ji}$, the dependence between i and j is symmetric. Hence, the null hypothesis can be designed as $H_0$: $d_{ij} = 0$, where $d_{ij} = r_{ij} - r_{ji}$. Remark that $d_{ij}$ is also a function of $\mathbf{H}$ with continuous first-order preference derivatives, which means that delta method is still applicable. Therefore, we can construct the statistic for test as $\sqrt{[\frac{T}{2}]}(\hat{d}_{ij} - d_{ij})$ and its asymptotic distribution is,

$$\sqrt{[\tfrac{T}{2}]}(\hat{d}_{ij} - d_{ij}) \xrightarrow{d} N(0, \nabla_{\mathrm{d}}(\mathbf{H})\mathbf{\Sigma}\nabla_{\mathrm{d}}^{\mathrm{T}}(\mathbf{H})). \tag{2-28}$$

By the way, we can also use the equal null hypothesis, $H_0$: $d_{ij}^2 = 0$. Then, under null hypothesis, the relative statistic is $\left(\frac{(\hat{d}_{ij} - d_{ij})}{\nabla_{\mathrm{d}}(\mathbf{H})\mathbf{\Sigma}\nabla_{\mathrm{d}}^{\mathrm{T}}(\mathbf{H})/\sqrt{[\frac{T}{2}]}}\right)^2$ whose asymptotic distribution is $\chi^2(1)$.

## *2.5 Asymmetry versus Variance*

Except for entropy, variance is another statistic frequently used to reflect the uncertainty of random vector (variable). In this section, we discuss relation between variance and asymmetry under the following conditions for any random vector $X=(X_1, \ldots, X_d) \in \mathbf{R}^d$:

C1. $\|E(X^{\mathrm{T}}X)\|_\infty < +\infty$, where $\|\bullet\|_\infty$ indicates the infinite norm of any real matrix.

C2. Suppose $f$ is the density function of $X$, then it satisfies two properties: (a) $f$ is positive and bounded, and (b) $\log f$ has a continuous second preference derivative.

Hereby, under C1 and C2, we have

$$\log f(x) = \log f(\mu) + \left(\frac{\partial \log f(x)}{\partial x}\right)^{\mathrm{T}}_{x=\mu} (x-\mu) + (x-\mu)^{\mathrm{T}} \left(\frac{\partial^2 \log f(x)}{\partial x \partial x^{\mathrm{T}}}\right)^{\mathrm{T}}_{x=\xi} (x-\mu) \quad (2\text{-}28)$$

where $\mu$ denotes the mean of $X$ and $\frac{\partial \log f(x)}{\partial x}$ is a transposition of gradient. Besides, $\xi = \lambda x + (1-\lambda)\mu$, where $\lambda \in (0,1)$. By setting $\left(\frac{\partial \log f(x)}{\partial x}\right)^{\mathrm{T}}_{x=\mu}$ and $\left(\frac{\partial^2 \log f(x)}{\partial x \partial x^{\mathrm{T}}}\right)^{\mathrm{T}}_{x=\xi}$ as A and $B(\lambda)$ respectively, we can easily obtain the following equation,

$$H(X) = -\{\log f(\mu) + E[(x-\mu)^{\mathrm{T}} B(\lambda)(x-\mu)]\}. \quad (2\text{-}29)$$

Meanwhile, under C2, we can also obtain the result below immediately,

$$H(X) = \lim_{\lambda \to 0} H(X) = -\left\{\log f(\mu) + \sum_{i=1}^{d} b_{ii} Var(X_i) + 2 \sum_{1 \le i < j \le d} b_{ij} Cov(X_i, X_j)\right\}, \quad (2\text{-}30)$$

where $b_{ij}$ is the element of matrix $B(\mu)$ at the *i-th* row and the *j-th* column. Particularly, when $d$=1, the relation above can be written as follow,

$$H(X) = -\{\log f(\mu) + B(\mu) Var(X)\}, \quad (2\text{-}31)$$

and $B(\mu)$ here is a scalar equaling to $\left(\log f(x)\right)^{(2)}\Big|_{x=\mu}$. Therefore, entropy somehow can be interpreted as a linear transformation of variance (or variance matrix).

Furthermore, recall that, for any two random variables (vectors) X and Y, our method is actually controlled by the values of H(X) and H(Y) respectively. Because once X and Y are given, mutual information $I_{XY}$ is fixed. Hence, we here discuss a theoretical phenomenon which might be a little anti-intuition, where linear correlation still causes significant asymmetrical dependence.

When the correlation coefficient $\rho_{XY}$=1, we have $Y$=**A**$X$+**b**, where **A** is a $d \times d$ matrix or 1-dimensional scalar and **b** can be a vector or scalar. To simplify the discussion, here we only discuss the case when $Y$ and $X$ are 1-dimensional random variables while this argument works the same to random vector. Assume $Y=aX+b$, where $a$ and $b$ are both non-zero constants. Based on (2-31), we have the following result,

$$\begin{aligned} H(Y) &= -\log f_Y(\mu_Y) - (\log f_Y(\mu_Y))^{(2)} Var(Y) \\ &= -\{\log f_Y(\mu_Y) + a^2 (\log f_Y(\mu_Y))^{(2)} Var(X)\} \end{aligned}. \qquad (2\text{-}32)$$

To demonstrate this relation clearly, we add another assumption—$X$ and $Y$ are from an identical distribution. Then, (2-32) is transformed into the following form,

$$H(Y) - H(X) = (1-a^2)(\log f(\mu))^{(2)} \sigma^2, \qquad (2\text{-}33)$$

where $\sigma^2=Var(X)=Var(Y)$. Take this back to (2-1), we can easily discover the difference of asymmetrical dependence between $X$ and $Y$ is quite significant once $|a|$ is sufficiently small or large. This discovery has been proved by our simulation work demonstrated in section 3.1 (experiment 1).

## 3. Simulation

The following two experiments serve as a demonstration of the applicability of the method proposed in section 2. In the first experiment(experiment 1), we introduce a micro-economical model, price equilibrium model between two firms, to exhibit that even absolutely linear transformation still engenders obvious asymmetrical dependence. More generally, in experiment 2, we display asymmetrical dependence caused by several different transformations between two random variables.

### *3.1 Experiment 1*

Suppose there are only two companies in a market named as enterprise 1 and 2. The commodities supplied by them are substitute to each other, though heterogeneity of product exists. Assuming there is no marginal but fixed cost, based on linear demand function, we introduce the following model,

$$\begin{cases} Q_1 = 24 - 4P_1 + 2P_2 \\ Q_2 = 24 - 4P_2 + 2P_1 \end{cases},$$

where $Q_i$ denotes the quantity of demand and $P_i$ indicates the price, i=1,2. Since we only need to concern fixed cost, for enterprise 1, total income (set as $\pi_1$) can be expressed as follow,

$$\pi_1 = P_1 Q_1 - C_1, \tag{3-1}$$

where $C_1$ means the fixed cost. Thus, reaction function of $P_1$ conditional of $P_2$ is,

$$P_1 = 3 + \frac{1}{4} P_2. \tag{3-2}$$

Furthermore, we assume $P_2$ is a random variable with probability space $(R^+, \mathcal{B}_{R^+}, \mu_2)$, where $\mu$ denotes probabilistic measure and $\mathcal{B}$ is a Borel field on $R^+$. Then, $P_1$ is also a random variable since it is a linear transformation of $P_2$. According to this structure, by setting the density of $P_2$ as $\frac{f_2(y)}{\mu_2(Y>0)}$, where $f_2(y)$ is the density function of normal distribution with mean 1 and variance 0.25, we generate stationary 5-dependent samples under different sample sizes (T=300, 500, 1000, 1500 and 2000).

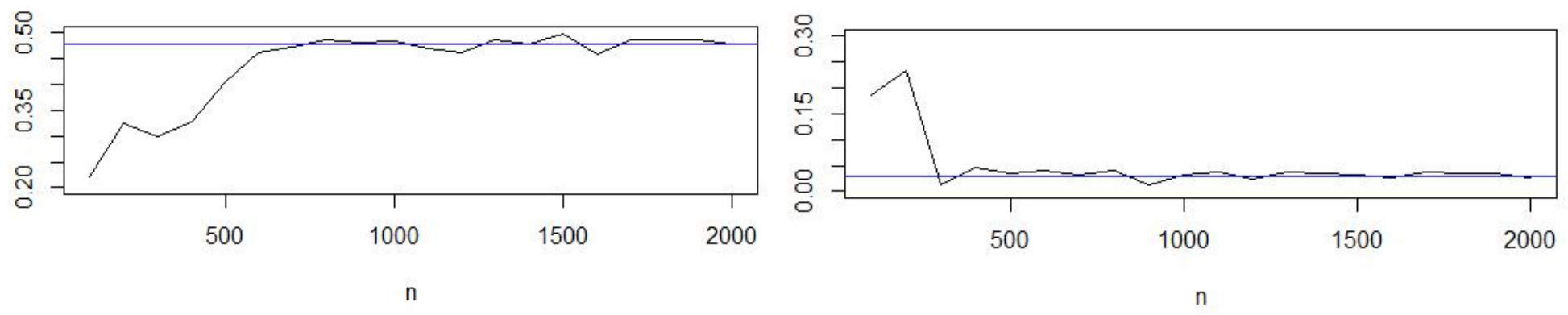

(a) estimation of $r_{21}$ (b) estimation of $r_{12}$

Figure 2. Performances of two estimators. (a) and (b) demonstrate the statistical performance of the estimation of $r_{21}$ and $r_{12}$ respectively under different sample size, where axis y represents the probable value that relative estimator takes.

Lines in (a) and (b) from Figure 2 indicate the true values calculated by Monte-carlo integration. Apparently, the absolute difference between $r_{12}$ and $r_{21}$ is approximately equal to 0.44. We could account for this phenomenon conceptually by comparing the variance of $P_1$ and $P_2$. Please remark that variance of $P_2$ is 16 times as large as that of $P_1$, which means the uncertainty of $P_2$ is far stronger. Hence, entropy, as a measure of uncertainty, will reflect that， shown by the significant asymmetries of dependence between $P_1$ and $P_2$.

### *3.2 Experiment 2*

According to (3-2), we generalize our discussion in experiment 1. Assuming that there are 6 individuals$\{X, X_2, X_3, \dots, X_6\}$ and all $X_i$s (i>1) are functions of $X$ as follow, where $X \sim N(0,1)$, $X_2 = 12/X$, $X_3 = e^X$, $X_4 = X^2$, $X_5 = 12/X^2$, $X_6 = Xe^X$.

We investigate the asymmetrical spatial association by estimating $r_{xi}$ and $r_{ix}$, i=2,...,5,6, under stationary 5 dependent samples with different sample sizes (T=500,1000,1500 and 2000). Similar to experiment 1, Figure 3 presents the statistical performance of estimator under different transformations.

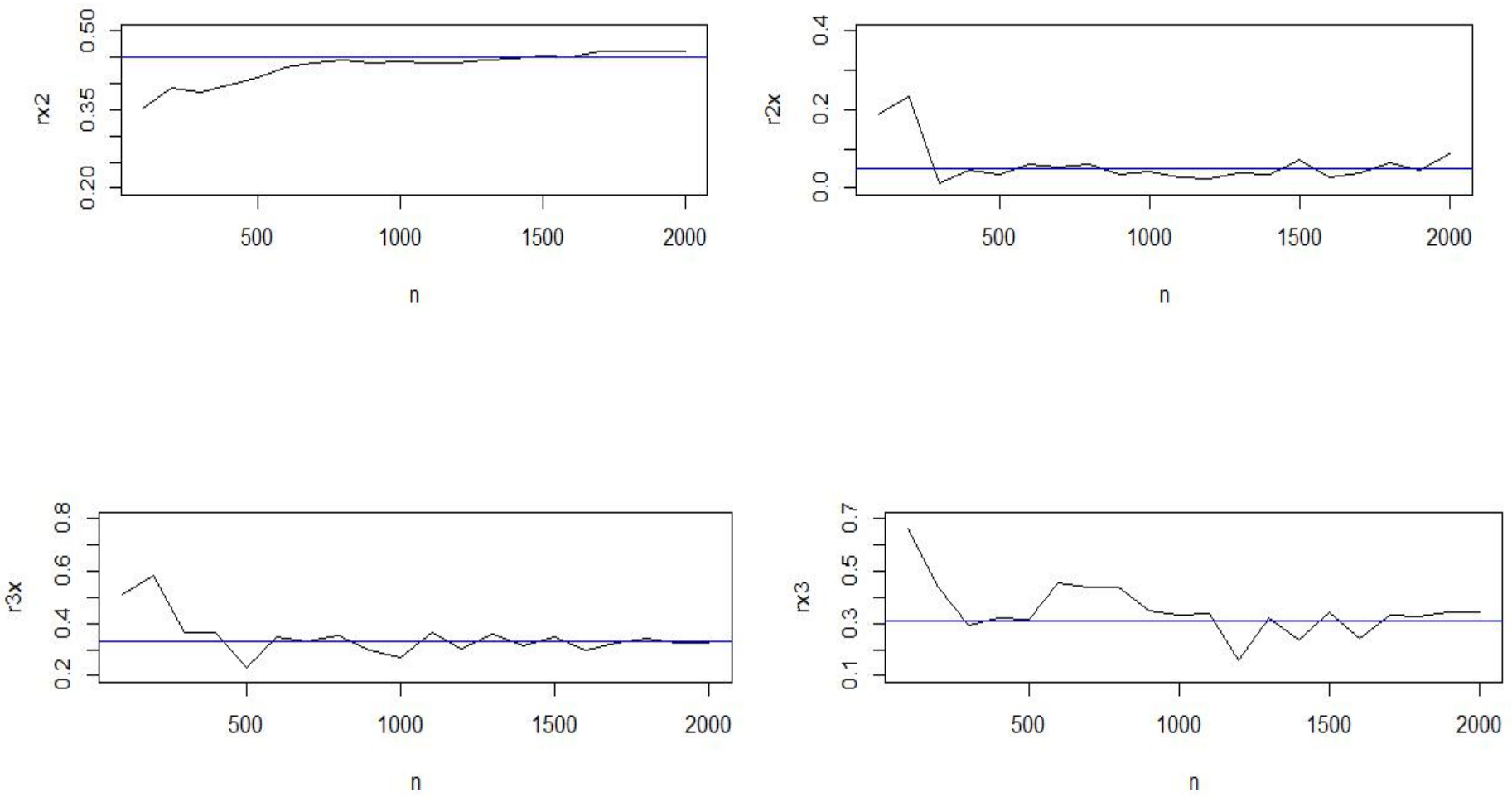

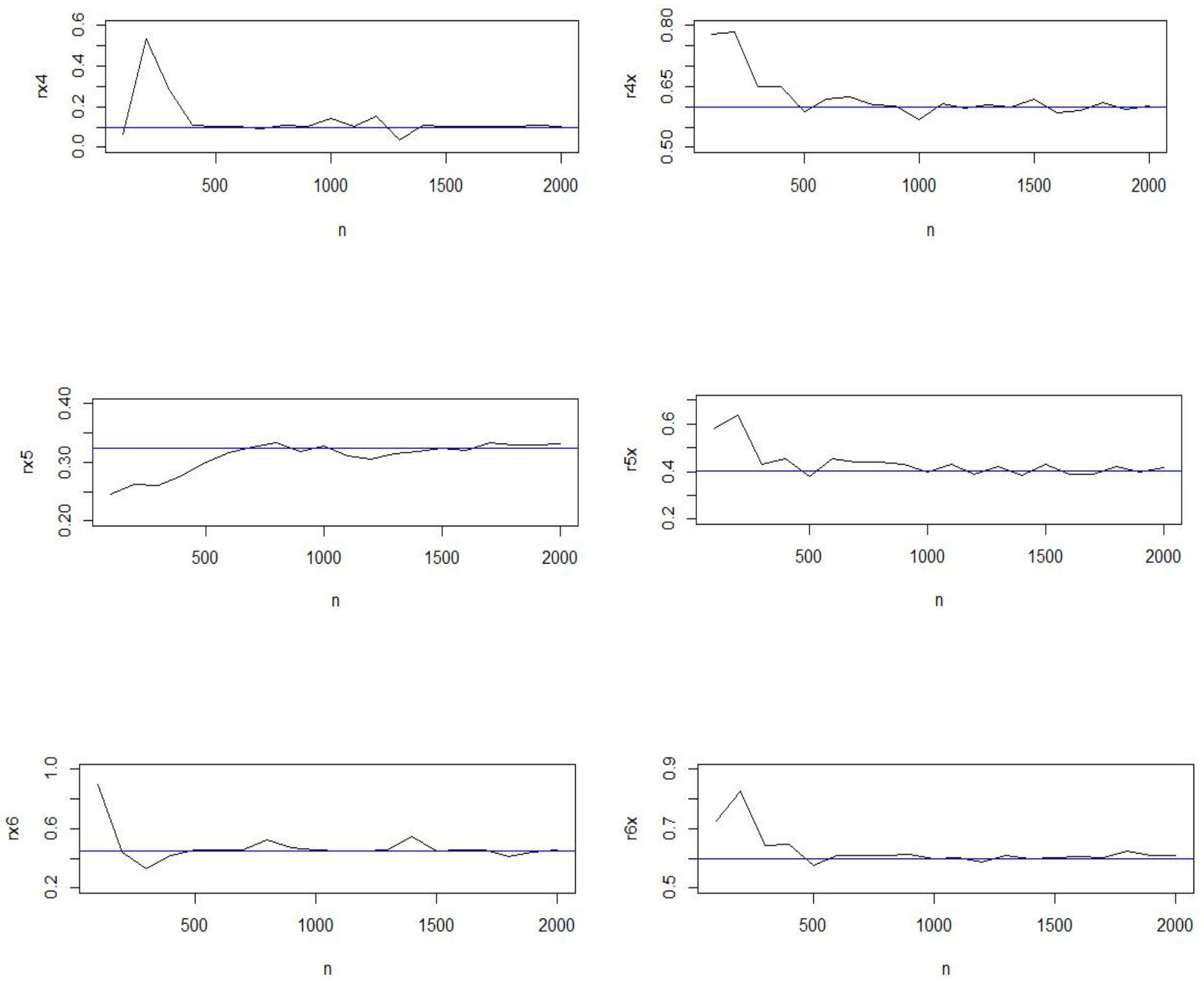

Figure 3. The statistical performances of different estimators. $r_{iX}$ and $r_{Xi}$ represent the asymmetrical dependences between $X$ and $X_i$.

These figures demonstrate the performance of estimators $r_{Xi}$ and $r_{iX}$,i=2,3,...,6. Axis y denotes the value of estimation, while the lines indicate the true values calculated by Monte-Carlo integration. Additionally, all of the deviation of estimators tend towards vanishing when T>500.

## 4. Conclusion

In this paper, we build an entropy-based method to measure and test the spatial association under heterogeneous environment. This method allows us to open the inner structure of spatial dependence and precisely investigate “local” and “asymmetrical” association for each observed individual without weighted matrix. Meanwhile, assumptions about the stationarity and dependency of spatio-temporal process are relatively general, which indicates that this method can theoretically fit most of the situation. Furthermore, if we can generalize the assumption of data over time, which is stationary m-dependent in this paper, the more suitability will our method have, since there is no assumption about data over space. On the other hand, though, in this paper, we only discuss a spatio-temporal process with continuous state space, this idea is still applicable for discrete state space.

However, there are still some disadvantages. Complexity of calculation of this method is relatively large, since given N individuals, we need to calculate N(N-1) times to fully gain the whole results. Besides, by using kernel-based entropy estimator, dimensionality curse still exists.

**Supplementary Materials**

**Proof of Lemma 1**

Here we only demonstrate the proof of (2-8), since proof of (2-7) has been well demonstrated by any nonparametric econometrics text book. Meanwhile, it is simple to tell that (2-9) is a direct extension of (2-7) and (2-8).

$$\because Var(\hat{f}_{12}(x_1,x_2)) = \frac{1}{([\frac{T+1}{2}]h_1h_2)^2} Var\left(\sum_{t=1}^{[\frac{T+1}{2}]} K_1\left(\frac{Z_{1t}-x_1}{h_1}\right)K_2\left(\frac{Z_{2t}-x_2}{h_2}\right)\right),$$

and by setting $K_1\left(\frac{Z_{1t}-x_1}{h_1}\right)K_2\left(\frac{Z_{2t}-x_2}{h_2}\right)$ as $K\left(\frac{Z_t-x}{h}\right)$, where x and $Z_t$ here denote $(x_1,x_2)$ and $(Z_{1t},Z_{2t})$ respectively, we directly have

$$Var(\hat{f}_{12}(x_1,x_2)) = \frac{1}{([\frac{T+1}{2}]h_1h_2)^2}\left[\sum_{t=1}^{[\frac{T+1}{2}]} Var(K\left(\frac{Z_t-x}{h}\right) + \sum_{|t-s|\le[\frac{m}{2}]} Cov\left(K\left(\frac{Z_t-x}{h}\right), K\left(\frac{Z_s-x}{h}\right)\right)\right]$$

Note that once a random sequence $\{X_i | i \in N^+\}$ is stationary m-dependent, its odd or even sub-sequence is $[\frac{m}{2}]$-dependent. Therefore, under AS.2, the term of covariance has the following upper bound,

$$\sum_{|t-s|\le[\frac{m}{2}]} Cov\left(K\left(\frac{Z_t-x}{h}\right),K\left(\frac{Z_s-x}{h}\right)\right)$$
$$=2\sum_{t=1}^{[\frac{T+1}{2}]}\sum_{0<s-t\le[\frac{m}{2}]} Cov\left(K\left(\frac{Z_t-x}{h}\right),K\left(\frac{Z_s-x}{h}\right)\right)$$
$$\le 2\sum_{t=1}^{[\frac{T+1}{2}]}\sum_{0<s-t\le[\frac{m}{2}]} \rho([\tfrac{m}{2}])Var\left(K\left(\frac{Z_t-x}{h}\right)\right)$$
$$=2[\tfrac{m}{2}]\rho([\tfrac{m}{2}])\sum_{t=1}^{[\frac{T+1}{2}]} Var\left(K\left(\frac{Z_t-x}{h}\right)\right)$$

Then, based on assumption of stationarity, we obtain

$$Var(\hat{f}_{12}(x))\le\left(\frac{2[\frac{m}{2}]\rho([\frac{m}{2}])+1}{[\frac{T+1}{2}]h_1h_2}\right)\left(f_{12}(x)\int\prod_{l=1}^{2}K_l^2(v)dv+O(\sum_{l=1}^{2}h_l^2+1)\right).$$

**Proof of lemma 2**

Here we mainly discuss (2-14) and (2-15). Firstly, for (2-14), according to stationarity, we have,

$$-E(\hat{H}_{12})=E\left(\frac{1}{[\frac{T}{2}]}\sum_{t=1}^{[\frac{T}{2}]}\log\hat{f}_{12}(x)\right)=E(\log\hat{f}_{12}(x)).$$

On the basis of the definition of expectation,

$$E(\log\hat{f}_{12}(x))=\int\log\left(\frac{1}{[\frac{T+1}{2}]}\sum_{t=1}^{[\frac{T+1}{2}]}K\left(\frac{Z_t-x}{h}\right)\right)f_{12}(x)dx$$

It is obvious that for any density function, the integral on its support set is equal to 1. Hence, due to this property we can rewrite the expectation above as follow,

$$E(\log\hat{f}_{12}(x))=\int f_{12}(z)dz\int\log\left(\frac{1}{[\frac{T+1}{2}]}\sum_{t=1}^{[\frac{T+1}{2}]}K\left(\frac{Z_t-x}{h}\right)\right)f_{12}(x)dx$$
$$=\int\left\{\underbrace{\int f_{12}(z)\log\frac{1}{[\frac{T+1}{2}]}\sum_{t=1}^{[\frac{T+1}{2}]}K\left(\frac{Z_t-x}{h}\right)dz}_{(1)}\right\}f_{12}(x)dx\cdot$$

Note that (1) is actually a kernel density estimation of density $f_{12}(x)$ for any given x and we here write it as $\hat{f}_{12}(x)$ . Thus, the following results can be shown by applying Lagrange's intermediate value theorem,

$$\begin{aligned} E(\log \hat{f}_{12}(x)) &= \int \left\{ \int f_{12}(z)(\log f_{12}(x) + \zeta^{-1}(\hat{f}_{12}(x) - f_{12}(x))dz \right\} f_{12}(x)dx \\ &= \int f_{12}(x)(\log f_{12}(x)dx + \zeta^{-1} \int Bias(\hat{f}_{12}(x)) f_{12}(x)dx \end{aligned}$$

where $\zeta = \theta \hat{f}_{12}(x) + (1-\theta) f_{12}(x)$ . Then, under AS.2-AS.7 and (2-7), we immediately obtain

$$Bias(\hat{H}_{12}) = \frac{1}{2\zeta} \left[ \sum_{l=1}^{2} \left( \kappa_{2(l)} h_l \right)^2 \int f_{ll}''(x) f(x)dx \right] + O(\sum_{l=1}^{2} h_l^3).$$

Secondly, as for (2-15), we have

$$\begin{aligned} Var(\hat{H}_{12}) &= \frac{1}{[\frac{T}{2}]} Var\left( \sum_{t=1}^{[\frac{T}{2}]} \log \hat{f}_{12}(x_t) \right) \\ &= \frac{1}{[\frac{T}{2}]} \{ \sum_{t=1}^{[\frac{T}{2}]} Var(\log \hat{f}_{12}(x_t)) + \underbrace{\sum_{|t-s| \le [\frac{m}{2}]} Cov(\log \hat{f}_{12}(x_t), \log \hat{f}_{12}(x_s)) \}}_{(2)} \end{aligned}$$

Similar to proof of lemma 1, according to relative assumptions, we have

$$(2) \le 2[\tfrac{m}{2}] \rho([\tfrac{m}{2}]) \sum_{t=1}^{[\frac{T}{2}]} Var\left( \log \hat{f}_{12}(x) \right).$$

Meanwhile, by using the same technique in the proof of (2-14), we have

$$\begin{aligned} E(\log \hat{f}_{12}(x))^2 &= \int f_{12}(x) \left( \log \hat{f}_{12}(x) \right)^2 dx \\ &= \int \left( \int \left( \log \hat{f}_{12}(x) \right)^2 f_{12}(z)dz \right) f_{12}(x)dx \\ &= \int \left( \int \left( \log f_{12}(x) \right)^2 + 2\zeta^{-1} \log \zeta (\hat{f}_{12}(x) - f_{12}(x)) f_{12}(z)dz \right) f_{12}(x)dx \\ &= \int \left( \int \left( \log f_{12}(x) \right)^2 + 2\zeta^{-1} \log \zeta (Bias(\hat{f}_{12}(x))) f_{12}(z)dz \right) f_{12}(x)dx \\ &= E(\log f_{12}(x))^2 + \kappa_2^2 \zeta^{-1} \log \zeta \sum_{l=1}^{2} \left( h_l^2 \int f_{ll}''(x) f_{12}(x)dx \right) \\ &= E(\log f_{12}(x))^2 + O(\sum_{l=1}^{2} h_l^2) \end{aligned}$$

Therefore, combining with results from proof of (2-14), we obtain

$$Var(\log \hat{f}_{12}(x)) = Var(\log f_{12}(x)) + o(1).$$

Take this result back, under assumptions of stationarity and limited variance, we instantly have the following result,

$$\begin{aligned} Var(\hat{H}_{12}) &\leq \frac{1}{[\frac{T}{2}]} Var(\log f_{12}(x)) + \frac{2[\frac{m}{2}]\rho([\frac{m}{2}])}{[\frac{T}{2}]} \sum_{t=1}^{[\frac{T}{2}]} Var\left(\log f_{12}(x)\right) \\ &= \frac{1}{[\frac{T}{2}]} Var(\log f_{12}(x)) + o(1) \end{aligned}$$

Thus, we finish the proof.

**Proof of lemma 3 and 4**

Apparently, content of lemma 3 is essentially same as lemma 4. Besides proofs of (2-22) and (2-23) are almost repetitive. Thus we here only demonstrate the proof of (2-23) in lemma 4. Based on definition of covariance, we directly have,

$$Cov(\hat{H}_i, \hat{H}_{ij}) = \frac{1}{[\frac{T}{2}]^2} \sum_{t,s} Cov(\log \hat{f}_i(x_{it}), \log \hat{f}_{ij}(x_s)),\ x_s = (x_{is}, x_{js}).$$

A fact which should not be ignored is that for any $i$, we have assumed $\{X_i(t)\}$ as a stationary sequence. Considering that $\log(.)$ is a monotonous and continuous mapping, it is obvious that for any i or (i,j), both $\{\log \hat{f}_i(x_{it})\}$ and $\{\log \hat{f}_{ij}(x_t)\}$ are stationary sequences as well, which means

$$Cov(\hat{H}_i, \hat{H}_{ij}) = Cov(\log \hat{f}_i(x_i), \log \hat{f}_{ij}(x)).$$

Therefore, we have

$$Cov(\hat{H}_i, \hat{H}_{ij}) = \underbrace{E(\log \hat{f}_i(x_i) \log \hat{f}_{ij}(x))}_{(a)} - \underbrace{E(\log \hat{f}_i(x_i)) E(\log \hat{f}_{ij}(x))}_{(b)}.$$

Due to (2-14) and (2-18), we can directly capture the asymptotic properties of (b) as follow,

$$\begin{aligned} (b) &= \left(E(\log f_i(x_i)) + O(h_i^2)\right)\left(E(\log f_{ij}(x)) + O(h_i^2 + h_j^2)\right) \\ &= E(\log f_i(x_i)) E(\log f_{ij}(x)) + O(h_i^2 + h_j^2) \end{aligned}$$

As for (a), we use the technique similar to the proof of lemma 2 to investigate it. Afterwards, we have

$$(a) = \int \left(\log \hat{f}_i(x_i) \log \hat{f}_{ij}(x)\right) f_{ij}(x) dx$$

$= \int \left( \int \int \log \hat{f}_i(x_i) \log \hat{f}_{ij}(x) f_i(z_i) f_{ij}(z) dz_i dz \right) f_{ij}(x) dx$, where $z = (z_i, z_j)$.

Meanwhile, by using first-order Taylor's expansion at $\hat{f}_i(x_i) = f_i(x_i)$ and $\hat{f}_{ij}(x) = f_{ij}(x)$, it has

$$\begin{aligned} (a) = E(\log f_i(x_i) \log f_{ij}(x)) + \eta_i^{-1} \log \zeta \int Bias(\hat{f}_i(x)) f_{ij}(x) dx \\ + \zeta^{-1} \log \eta_j \int Bias(\hat{f}_{ij}(x)) f_{ij}(x) dx \end{aligned}.$$

Then, via lemma 1 and corollary 1, we complete the proof.

**Proof of Theorem 2**

$$\begin{aligned} \hat{r}_{ij} - r_{ij} &= \frac{\hat{H}_j - \hat{H}_{ij}}{\hat{H}_i} - \frac{H_j - H_{ij}}{H_i} \\ &= \frac{H_i \hat{H}_j - H_i \hat{H}_{ij} - \hat{H}_i H_j + \hat{H}_i H_{ij}}{\hat{H}_i H_i} \\ &= \frac{(H_i \hat{H}_j - \hat{H}_i H_j) - (H_i \hat{H}_{ij} - \hat{H}_i H_{ij})}{\hat{H}_i H_i} \\ &= \frac{H_i (\hat{H}_j - H_j) - H_i (\hat{H}_{ij} - H_{ij}) + (\hat{H}_i - H_i)(H_{ij} + H_j)}{\hat{H}_i H_i} \end{aligned}$$

Hence, in the light of (2-13) and (2-17), (2-24) is proved. For (2-25), we use second-order Taylor's expansion on $\hat{r}_{ij}$ at $\hat{\mathbf{H}} = \mathbf{H}$ and calculate the expectation on both sides. Under lemma 2, corollary 2 and lemma 3, we can finish this proof.

**Proof of Lemma 5**

Please remark

$$\begin{aligned} \sqrt{[\tfrac{T}{2}]}(\hat{\mathbf{H}}^{\mathrm{T}} - \mathbf{H}^{\mathrm{T}}) &= \sqrt{[\tfrac{T}{2}]}(\hat{\mathbf{H}}^{\mathrm{T}} - E(\hat{\mathbf{H}}^{\mathrm{T}})) + \sqrt{[\tfrac{T}{2}]}(E(\hat{\mathbf{H}}^{\mathrm{T}}) - \mathbf{H}^{\mathrm{T}}) \\ &= \underbrace{\sqrt{[\tfrac{T}{2}]}(\hat{\mathbf{H}}^{\mathrm{T}} - E(\hat{\mathbf{H}}^{\mathrm{T}}))}_{(c)} + \underbrace{\sqrt{[\tfrac{T}{2}]} Bias(\hat{\mathbf{H}}^{\mathrm{T}})}_{(d)} \end{aligned}.$$

According to lemma 2, corollary 2, lemma 4 and AS.7, the second term (*d*) of the second equation expressed above is equivalent to $o(1)$, i.e., $(d) = o(1)$, which means

$$\lim_{T\to+\infty}\sqrt{[\tfrac{T}{2}]}(\hat{\mathbf{H}}^{\mathrm{T}} - \mathbf{H}^{\mathrm{T}}) = \lim_{T\to+\infty}\sqrt{[\tfrac{T}{2}]}(\hat{\mathbf{H}}^{\mathrm{T}} - E(\hat{\mathbf{H}}^{\mathrm{T}})).$$

In terms of the central limit theorem under stationary m-dependent sequence, proof is completed.

**Connection between copula and mutual information**

All of the discussion here is based on Sklar(1959). Consider two random variables X and Y and a copula function C(U,V), where $U = F_X(x)$, $V = F_V(v)$ are uniform marginal distributions. Mathematically, the copula density can be written as:

$$c(u,v) = \frac{\partial}{\partial F_y}\left(\frac{\partial C}{\partial F_x}\right) = \frac{\partial^2 C(u,v)}{\partial F_X \partial F_y}.$$

Note that if we transform the original variables via their CDF functions, then the joint density function of transformed variables U and V are actually the copula density c(u,v). We know that any joint distribution can be written in the terms of copulas; i.e,

$$C(u,v) = F(x,y).$$

Taking second-order partial derivatives with respect to x and y respectively, on both sides yields:

$$\frac{\partial C}{\partial y}\left(\frac{\partial C(u,v)}{\partial x}\right) = f(x,y),$$

$$\frac{\partial C^2(u,v)}{\partial F_x \partial F_y}\frac{\partial F_x \partial F_y}{\partial x \partial y} = f(x,y),$$

$$c(u,v)f(x)f(y) = f(x,y),$$

$$\log c(u,v) = \log\frac{f(x,y)}{f(x)f(y)}.$$

The last equation indicates the critical connection between mutual information.